\documentstyle[12pt]{article}

\makeatletter
\@addtoreset{equation}{section}

\makeatother

\renewcommand{\baselinestretch}{1.5}

\def\beq{\begin{equation}}
\def\eeq{\end{equation}}
\def\bea{\begin{eqnarray}}
\def\nn{\nonumber \\ }
\def\eea{\end{eqnarray}}
\def\vs{\vspace{2mm}}
\def\ds{\displaystyle}

\def\ni{\noindent}

\def\req#1{(\ref{#1})}
\def\ie{{\it i.e.}\ }
\def\eg{{\it e.g.}\ }
\def\rep{{\rm Re}\ }
\def\imp{{\rm Im}\ }
\def\slashed#1{\slash\!\!\!\!#1}

\begin{document}

\begin{center}

{\LARGE \bf Dynamical Generation of Spacetime Signature by Massive Quantum
Fields on a Topologically Non-Trivial Background}

\vskip4ex

\renewcommand\baselinestretch{0.8}

Sergei D. Odintsov\footnote{
on leave of absence from Dept of Mathematics and Physics,
Pedagogical Institute, 634041 Tomsk, Russian Federation,
E-mail: odintsov@ebubecm1.bitnet},
August Romeo, \\
{\it Dept ECM, \\
Faculty of Physics, University of Barcelona, \\
Diagonal 647, 08028 Barcelona, Catalonia, Spain} \\
\vspace{0.5cm}
and \\
Robin W. Tucker\footnote{E-mail rwt@lavu.physics.lancaster.ac.uk}, \\
{\it School of Physics and Materials, \\
University of Lancaster, Bailrigg, Lancs. LA1 4YB, UK}

% to obtain the figure, contact the first author %%%%%%%%%%%%%%%%
\renewcommand\baselinestretch{1.4}

\vspace{1cm}

{\bf Abstract}

\end{center}

The effective potential for a dynamical Wick field (dynamical signature)
induced by the quantum effects of massive fields on a topologically
non-trivial $D$ dimensional background is considered.
It is shown that
when the radius of the compactified
dimension is very small compared with $\Lambda^{1/2}$ (where $\Lambda$
is a proper-time cutoff), a flat metric with Lorentzian signature
 is preferred on ${\bf R}^4 \times {\bf S}^1$.
When the compactification radius becomes larger a careful analysis of the
1-loop effective potential indicates that a Lorentzian signature is
preferred in both $D=6$ and $D=4$ and that these results are relatively
stable under metrical perturbations.

\newpage

\section{Introduction}

In classical physics, the principle of causality
is traditionally required as a basic assumption.
According to relativistic notions,
it may be expressed in terms of a Lorentzian geometry for a
4-dimensional differentiable spacetime
manifold possessing a continuous field of light cones with a
preferred time orientation.
However, when applying quantum theory to spacetime itself
\cite{DiracCJM}-\cite{BIPRD},\cite{HarHawPRD}-\cite{HalHarPRD},
\cite{KundtCMP}-\cite{Brill}
a number of intriguing
possibilities emerge that force one to re-examine the pre-eminent role of a
Lorentzian geometry as the arena for physical phenomena on all scales.

In a quantum description that includes gravitation,
it is natural to consider semi-classical 4-geometries that undergo a change of
metric signature together with a possible {\it topological transition}
\cite{HawMosPLB}-\cite{MosPolNPB}.
It has long been conjectured that classical
spacetime may have an intricate topology on
a microscopic scale or during a Planckian era,  and there have recently
been attempts to examine the
consequences of such a structure on macroscopic physics, including the
possibility that the constants of nature may have their origin in the
topology of spacetime \cite{ColNPB}.
  A number of
 cosmological models also envisage a primordial geometry that undergoes a
signature transition from Euclidean to Lorentzian thereby separating the
origin of classical time from the initial quantum creation of the
Universe.

The phenomenon of dynamical spacetime
topology
change
 may be  accompanied by a dynamical signature change of
the spacetime metric. Thus, a more general formulation of
gravitation should accommodate geometries with degenerate metrics and
non-trivial topologies.
In the last year or so,
there has been considerable activity in the analysis of
classical solutions to Einstein's gravitational field
equations that admit degenerate metrics
\cite{DTCQG}-\cite{HaywardCQG}.
Their relevance
 to a number of approaches to the full quantum theory has also
been pursued.

The existence  of certain
manifolds that cannot sustain a global Lorentzian
metric implies that  signature is inherently dynamical in
theories with dynamical topology.
The question is how best to model such effects in the language of quantum
field theory.

Percacci \cite{PercacciNPB}
seems to have been the first to offer a formalism in which one can
discuss such notions at the level of effective actions.
 His approach is to dissociate the conventional
geometrical interrelations between the metric tensor components and a
field
of coframes, and work in close analogy with the Higg's model in
non-Abelian gauge theories.
Classical geometry is then regarded as an interpretation of
certain expectation values  which minimise an effective action. Greensite
\cite{gPLB,gcPre}
has developed this idea further by assuming that a particular pattern of
signatures arises dynamically as a result of a dynamical phase field
that interpolates between signatures. In particular, he assumes that the
effective action can become complex, thereby destroying all vestige of a
geometrical interpretation for the gravitational degrees of freedom
 in the theory. However if one
adopts  Percacci's viewpoint then this extension might be pursued at
the quantum level, since one should
only expect a spacetime geometry to emerge in
some classical limit. Greensite has argued that, at least for the
free  scalar field theory interacting with such a dynamical Wick field, the
Lorentzian signature of a four dimensional manifold can be predicted as a
ground state expectation value.   There are a number of issues raised by
this claim.
In particular, Elizalde et al \cite{eorCQG} have examined
the dependence of the result on  the
influence of other topological configurations on the computation. It
appears that the minimising signature is rather sensitive to the details of
the field system under consideration. Thus  the approach offers a means of
exploring self-consistent solutions to effective action theories that can
correlate the observed properties of classical gravitation with the quantum
structure of matter.

In the present paper we discuss  the effective potential of
a dynamical Wick field induced by the quantum effects of massive quantum fields
in
$D$-dimensions.
%non-Minkowskian spacetime.
In section 2, after a short review of the formalism of
refs. \cite{gPLB,gcPre}, we develop the general structure of the 1-loop
effective
potential for a dynamical Wick field.
In section 3 we discuss an expansion for this
potential  when topological effects are relevant,
i.e.  when the radius of the compactified dimension is very small
compared with $\Lambda^{1/2}$, where  $\Lambda$ is a proper-time cutoff.
Section 4 is devoted to study the case of
%the stability of these results.
large compactification radius, in addition to mass corrections, for
$D=4$ and $D=6$.
An appendix is included, where we outline the
application of the same type of calculation to the study of curvature
effects on massless fields in ${\bf R}^{D-N} \times {\bf S}^N$
spacetimes, with ${\bf S}^N$ meaning the
$N$-dimensional De Sitter space.

\section{Induced effective potential}

We describe  briefly the formalism employed in \cite{gPLB,gcPre}.
Consider the matrix
\beq
\eta_{ab}=\rm{diag}( e^{i\theta}, 1, 1, \dots, 1).
\label{etaab}\eeq

 We shall refer to this as a complexified
spacetime metric. It has a Euclidean signature for
$\theta=0$, and a Lorentzian one for $\theta=\pm\pi$.
It was suggested in \cite{gPLB} that the Wick angle $\theta$ in
\req{etaab} should be treated as a dynamical degree of freedom.
% in the domain  $\theta\in [-\pi, \pi]$.
In order to effect a Fourier analysis below we shall require that
$\cos\theta/2 > 0$ and we shall restrict our attention to values of $\theta\in[
-\pi, \pi]$.

We compute the 1-loop effective potential $V(\theta)$ as a function of
$\theta$ under the following assumptions \cite{gPLB,gcPre}:

Let
\beq
Z=\int d\mu(e, \phi, \psi, \bar\psi )
\exp\left[ -\int d^Dx \sqrt{g} {\cal L} \right],
\label{ZFPI}\eeq
with $e$ standing for the vielbein, $\phi$ for bosons and
$\psi, \bar\psi$ for fermions.
$g_{\mu \nu}=e^a_{\mu} \eta_{ab}e^b_{\nu}$, with the above
$\theta$-dependent $\eta_{ab}$.

%\item This is what was called assumption 2:

Following \cite{gPLB,gcPre}, we assume that
the integration measure for scalar fields is given by the
real-valued, invariant volume measure (DeWitt measure) in superspace
$d\mu(\phi)=D\phi\sqrt{ |G| }$, where $G$ is the determinant of the
scalar field supermetric $G(x,y)=\sqrt{g}\delta(x-y)$.

{}From \req{ZFPI} and under these conditions,
the one-loop potential induced by a
free bosonic scalar field of mass $m_B$ in flat spacetime
%(\ie $g_{\mu \nu}=e^a_{\mu} \eta_{ab} e^b_{\mu}=\eta_{\mu \nu}$)
is given by
\beq
%V_m(\theta )=
-{ \log \det^{-1/2}[
-\sqrt{\eta}( \eta^{ab} \partial_a \partial_b -m_B^2) ] \over
\int d^Dx } .
\label{Vlogdet}\eeq
We work in the representation where
%fermion fields have four
%times the degrees of freedom of scalars.
the  contribution from every free fermion of
mass $m_F$  (neglecting terms proportional to
$ \ds \cdot { \log \det[ \eta ] \over \int d^Dx}$
\footnote{the exponential of such terms is a factor which can be
absorbed in our integration measure.}
) is
$-\ds { \log \det[ \slashed{D}_{m_F} ] \over \int d^Dx }$,
where $\slashed{D}_{m_F}=i\slashed\partial -m_F$.
Taking into account Dirac conjugacy,  one has
$\det[ i\slashed\partial -m_F ]= \det^{1/2} [ \partial^2 -m_F^2 ]$.
Therefore, up to the above mentioned irrelevant terms,
this contribution to the one-loop effective potential
%reads
%\beq -{ \log \det^{+1/2}[
%-\sqrt{\eta}( \eta^{ab} \partial_a \partial_b -m_F^2) ] \over
%\int d^Dx } .
%\eeq
behaves like \req{Vlogdet} but with the
determinant raised to the powerr  $+1/2$, instead of $-1/2$  and $m_B$
 replaced by $m_F$.
After taking the logarithm, this implies  a  relative sign between boson and
fermion contributions. As a result
we can write the complete one-loop effective potential $V(\theta)$ as
\beq V(\theta)=\ds\sum_B V_{m_B}(\theta) - \sum_F V_{m_F}(\theta), \eeq
where
\beq
V_m(\theta ) \equiv
{ {1 \over 2} \log \det[
-\sqrt{\eta}( \eta^{ab} \partial_a \partial_b -m^2) ] \over
\int d^Dx } .
\eeq

Use of heat-kernel regularization gives \cite{gPLB,gcPre}
\beq
V_m(\theta)=-{1 \over 2}\int_{\Lambda}^{\infty} {ds \over s}
\int {d^Dp \over (2\pi)^D} \
e^{\ds -s[ \alpha p_0^2 + \beta ( \vec{p}^2 + m^2 )]},
\label{hkregVm}\eeq
where $\Lambda$ is a proper-time cutoff,
$p=\{p_0,p_1,.....,p_{D-1}\}$ and
\beq
\alpha\equiv e^{-i {\theta \over 2}}, \hspace{2cm}
\beta\equiv e^{i {\theta \over 2}}.
\eeq
 It is explained in \cite{gcPre} why
heat-kernel regularization is to be preferred in this context.

The potential $V_m(\theta )$ is complex.
As discussed in \cite{gPLB,gcPre},
%the value of
%$\theta$ to be picked as the best solution according to this potential
%---which we shall call $\theta=\bar\theta$--- will have to satisfy
%the requirements
we seek a solution $\theta=\bar\theta$ that ensures:
\beq
\begin{array}{lr}
\imp V(\bar\theta)&\mbox{is stationary,} \\
\rep V(\bar\theta)&\mbox{is  a minimum.}
\end{array}
\label{requi}
\eeq

These conditions \req{requi} are essentially {\it ad hoc} and
can only be justified a posteriori.

In order to study topological effects, periodic compactifications
of one dimension  will be considered (as observed in \cite{eorCQG},
once ${\bf R}^{D-1} \times {\bf S}^1$ has been studied, it is
not difficult to  extend  results to
${\bf R}^{D-n} \times {\bf S}^n$.)
We shall take  $p_1$,  discretized with
\[ p_1^2={n^2 \over R^2}, \ n \in {\bf Z}. \]
where $R$ is some fundamental length.
Consequently, in the expression of $V_m$ we  let
\[ \int {d p_1 \over 2 \pi} \to
{1 \over 2 \pi R} \sum_{n=-\infty}^{\infty}, \]
and, integrating over the remaining $p$-components, we find
\beq
V_m(\theta)=
-{1 \over (4 \pi)^{D+1 \over 2} R \alpha^{1 \over 2} \beta^{D-2 \over 2}}
\int_{\Lambda}^{\infty} ds \ s^{-{D-1 \over 2}-1} \ e^{-\beta m^2 s} \
\theta_3\left( 0 \left\vert {s \beta \over \pi R^2} \right. \right) ,
\label{Vt3}\eeq
where
\beq
\theta_3(0|z) \equiv \sum_{n=-\infty}^{\infty} e^{-\pi z n^2}
\eeq
is a Jacobi theta function.

It is convenient to introduce the
dimensionless variables
\bea
x&\equiv&\ds\beta\Lambda m^2, \nn
y&\equiv&\ds{\beta\Lambda \over R^2}
\eea

and write
\beq
V_m(\theta)=
-{1 \over (4 \pi)^{D+1 \over 2} R \Lambda^{D-1 \over 2}}
e^{-i{ D-3 \over 4}\theta }
\left[
x^{D-1 \over 2}\Gamma\left( -{D-1 \over 2}, x \right)
+2\sum_{n=1}^{\infty}
(x+yn^2)^{D-1 \over 2}\Gamma\left( -{D-1 \over 2}, x+yn^2 \right)
\right] .
\label{Vgen}
\eeq

in terms of the incomplete gamma function
\beq
\Gamma(a,z) \equiv \int_z^{\infty} dt \ t^{a-1} \ e^{-t} .
\label{igf}\eeq

%This general ---but not yet too meaningful--- expression is the
%starting point for considering specific physical situations, in which
%its particular contents will be examined in detail.

It is this expression we analyse as a function of $\theta, x, y, m$.
Note our choice of the regularization scheme has made $V$ cutoff
dependent.

\section{Strong topological effect: small $R$ $(|y|\gg 1)$}

\subsection{Small radius and large mass
%$R \ll 1, m \gg 1$
$(|y|\gg 1, |x|\gg1)$}

We note that $|y|\gg 1$ means $\Lambda/R^2 \gg 1$, which amounts to
saying that $R$ is small compared with $\Lambda^{1/2}$.
Similarly, $|x|\gg 1 $ is equivalent to $m \gg 1/\Lambda^{1/2}$.
 The asymptotic expansion of the incomplete
gamma function \req{igf} with a large second argument is
\beq
x^a \Gamma( -a, x)  \sim
x^{-1} e^{-x} \left[ 1 +
\sum_{k=1}^{\infty} {\Gamma( -a) \over \Gamma(-a-k)}{1 \over x^k}
\right], |x| \gg 1.
\label{GaAs}\eeq
Note that, interpreting the quotient of ordinary gamma functions
as a finite product, this holds for both non-integer and integer $a$
(here $a=(D-1)/2$). Inserting this formula into \req{Vgen}, one gets
\beq
V_m(\theta)=
-{1 \over (4 \pi)^{D+1 \over 2} R \Lambda^{D-1 \over 2}}
e^{-i{ D-3 \over 4}\theta }
\left[ {e^{-x} \over x} \left( 1 +O\left( 1 \over x \right) \right)
+2\sum_{n=1}^{\infty} {e^{-(x+yn^2)} \over x+yn^2 }
\left( 1 +O\left( 1 \over x+yn^2 \right) \right) \right] ,
\eeq
from which we  extract the leading term
\beq
V_m(\theta)\approx
-{1 \over (4 \pi)^{D+1 \over 2} R \Lambda^{D+1 \over 2}m^2 } \
e^{\ds -\Lambda m^2 \cos{\theta \over 2}
-i \left[
{D-1 \over 4} \theta + \Lambda m^2 \sin{\theta \over 2}
\right] .
}
\label{Vplat}\eeq
Typical curves representing $\rep V(\theta)$ and $\imp V(\theta)$
have been plotted in Fig. 1.
This sort of potential was already studied by the authors in
\cite{eorCQG}, so that the comment made there applies also to the
present case. $V_m(\theta)$ drastically changes its nature depending
on the specific value of $\Lambda m^2$. When this quantity is small, the
real and imaginary
parts of $V_m(\theta)$ are approximately sinusoidal, but as it grows,
 they tend to form a
flat plateau around the origin (notice that our assumption in
this subsection implies that $\Lambda m^2$ must be large).
Since,  under these circumstances there is no coincidence
between the minima of $\rep V$ and the stationary points of $\imp V$,
we cannot speak of any preferred $\theta$.

\subsection{Small radius and small mass
%$R \ll 1, m \ll 1$
$(|y|\gg 1, |x|\ll1)$}

Now, in order to expand the first term in \req{Vgen}, we need a
power series for the same  gamma function when its second
argument, $x$, is small. The expansion  is
\beq
x^a \Gamma( -a, x) =
\left\{ \begin{array}{ll}
\ds x^a \Gamma(-a)
-\sum_{n=0}^{\infty} { (-1)^n \over (n-a) n! } x^n, &a \ne 0,1,2, \dots
\\
\ds x^a {(-1)^a \over a!} [ \psi(a+1) - \ln x ]
-\sum_{n=0 \atop n \ne a}^{\infty} { (-1)^n \over (n-a) n! } x^n,
&a = 0,1,2, \dots
\end{array} \right.
\label{GaPo}\eeq
An equivalent series was employed in \cite{gcPre}.
Note, however, that our expression for integer $a$ is handier than the
one used in that reference. In fact, we have obtained it from the
noninteger-$a$ case by analytic
continuation in $a$, and its particular forms for every value of this
parameter coincide with the ones resulting from \cite{gcPre}.
In any case, the power series obtained will be  denoted by
\beq
x^a \Gamma( -a, x) =
b_a^{(a)} x^a + b_{(L)}^{(a)} x^a \ln x +\sum_{n\ge 0} b_n^{(a)} x^n .
\label{bPo}\eeq
Since the values of interest are $a=(D-1)/2$, we list the
special forms corresponding to the first integer and half odd integer
$a$'s in Table \ref{Tb1}. In particular the fifth and seventh rows
will be used in the cases to be studied later.

\begin{table}[htb]
\begin{center}
\begin{tabular}{rl}
$a$&power series for $x^a \Gamma( -a, x)$ \\ \hline \vs
0&$\ds -\gamma-\ln x +x -{x^2 \over 4} +{x^3 \over 18} + O(x^4) $ \\ \vs
${1\over 2}$&$\ds 2-2\sqrt{\pi}x^{1/2}+2x-{x^2\over 3} +{x^3 \over 15}
+ O(x^4)$ \\ \vs
1&$\ds 1+(\gamma-1)x+x\ln x -{x^2 \over 2} +{x^3 \over 12} + O(x^4) $
\\ \vs
${3\over 2}$&$\ds {2 \over 3}-2x+{4 \over 3}\sqrt{\pi} x^{3/2} -x^2
+{x^3 \over 9} + O(x^4) $ \\ \vs
2&$\ds {1 \over 2}-x+{1 \over 2}\left( {3 \over 2}-\gamma \right) x^2
-{1\over 2} x^2 \ln x + {x^3 \over 6} + O(x^4) $ \\ \vs
${5\over 2}$&$\ds {2 \over 5}-{2 \over 3}x+x^2
-{8 \over 15}\sqrt{\pi}x^{5/2} +{x^3\over 3} + O(x^4) $\\ \vs
3&$\ds {1 \over 3}-{x \over 2}+{x^2 \over 2}
-{1 \over 6}\left( {11 \over 6 }-\gamma \right) x^3
+{1 \over 6} x^3 \ln x + O(x^4) $ \\
\end{tabular}
\end{center}
\caption{$x^a \Gamma( -a, x)$ as a power series in $x$.}
\label{Tb1}\end{table}

%Once the necessary expansion \req{bPo} has been explicitly found,
We
substitute \req{bPo} in  the first term  of \req{Vgen}, while the remaining
contribution is expressed in terms of the asymptotic expression \req{GaAs}
%because $|y|$ is still large. With this,
\beq
\begin{array}{ll}
\ds V_m(\theta)=
-{1 \over (4 \pi)^{D+1 \over 2} R \Lambda^{D-1 \over 2}}
e^{-i{ D-3 \over 4}\theta }&
\ds\left[ b_{D-1 \over 2}^{({D-1\over 2})} x^{D-1 \over 2}
+ b_{(L)}^{({D-1\over 2})} x^{D-1 \over 2} \ln x
+\sum_{n\ge 0} b_n^{({D-1\over 2})} x^n \right. \\
&\ds\left.
+2\sum_{n=1}^{\infty} {e^{-(x+yn^2)} \over x+yn^2 }
\left( 1 +O\left( 1 \over x+yn^2 \right) \right) \right] .
\end{array}
\eeq
When $x=0$ and $|y| \to \infty$, the only contribution left is the
leading term $b_0$ in the \req{bPo} series, which, for $D>1$ takes on
the value
\beq b_0={2 \over D-1}. \label{valb0}\eeq
Then
%, the remaining part of
the potential reads
\beq
V_m(\theta)\approx
-{2 \over (D-1) (4 \pi)^{D+1 \over 2} R \Lambda^{D-1 \over 2}}
\ e^{\ds -i{D-3 \over 4}\theta } .
\eeq
If several fermions and bosons are present, this contribution appears
multiplied by $-\Delta n\equiv -(n_F-n_B)$, where $n_F$ ($n_B$) is
number of fermions (bosons).
%Of course, this is what one obtains, when one space component is
%compactified
%and $R\to 0$, for {\it massless} particles \cite{eorCQG}.
In this limit one recovers the  result for massless particles found in
\cite{eorCQG}.
As reported
in that reference,
%leaving
%aside the special cases $n_F=n_B$ or $D=3$ ---in which the $\theta$
%dependence is lost---
the only solution satisfying the requirements \req{requi} is
$n_F>n_B$ and $\theta=\pm\pi, D=5$.
%Therefore, for small radius, the Lorentzian signature is chosen in
%$D=5$, opening a window to Kaluza-Klein theories.

\section{Weak topological effect: large $R$ $(|y|\ll 1)$}

Going back to the initial form of $V_m(\theta)$,
 we apply
%inside
%eq. \req{Vt3} itself
the Jacobi theta function identity
\beq
\theta_3(0|z)=
{1 \over \sqrt{z}}
\theta_3\left( 0 \left\vert {1 \over z} \right. \right) .
\label{recitheta3}
\eeq
to eq. \req{Vt3}.
The resulting integrals will be evaluated by conveniently splitting
the integration domains and using
\beq
\int_0^{\infty} ds \ s^{\nu-1} \ e^{-as-{b \over s}}
=2 \left( b \over a \right)^{\nu \over 2} K_{\nu}(2 \sqrt{ab} )
\eeq
(where $K_{\nu}$ is the modified Bessel function of index $\nu$),
together with \req{GaAs} in combination with the exponential power series.
We find, for each $n$,
\[
\int_{\Lambda}^{\infty} ds \ s^{-{D \over 2}-1}
e^{\ds -{\pi^2 R^2 n^2 \over \beta s}-\beta m^2 s } =
\]
\beq
2\left( \pi R n \over \beta m \right)^{-{D \over 2}}
K_{-{D \over 2}}(2 \pi n m R)
-{1 \over \Lambda^{D \over 2}}
\left( \pi^2 R^2 n^2 \over \beta \Lambda \right)^{-1}
e^{\ds -{\pi^2 R^2 n^2 \over \beta \Lambda}-\beta m^2 \Lambda }
\left[ 1
+O\left(
\left( \pi^2 R^2 n^2 \over \beta \Lambda \right)^{-1}
\right)
\right]
\eeq
The effective potential becomes
\beq
\begin{array}{ll}
\ds V_m(\theta)=
-{ \pi^{1 \over 2} \over
(4 \pi)^{D+1 \over 2} \alpha^{1 \over 2} \beta^{D-1 \over 2}
\Lambda^{D \over 2} }&
\ds\left[ x^{D \over 2} \Gamma\left( -{D \over 2}, x \right) \right. \\
&\ds +4 x^{D \over 2} \sum_{n=1}^{\infty}
\left( \pi n \sqrt{x \over y} \right)^{-{D \over 2}}
K_{-{D \over 2}}\left( 2 \pi n \sqrt{x \over y} \right) \\
&\ds\left. -2 \sum_{n=1}^{\infty} {y \over \pi^2 n^2}
e^{ -{\pi^2 n^2 \over y} -x}
\left( 1 + O\left( y \over \pi^2 n^2 \right) \right)
\right] .
\end{array}
\label{Vgen2}\eeq
We discuss this for various values of $|x|$ and $|y|$.

\subsection{Large radius and large mass
%$R \gg 1, m \gg 1$
$(|y|\ll 1, |x|\gg 1)$}
We consider the domain where
\[ \left\vert {x \over y} \right\vert \gg 1. \]
%We start by working out the above Bessel function $n$-series in order to
%obtain
We need an expression for $V_m$ asymptotically valid for large
values of $a\equiv\sqrt{x/y}$.
To this end, we take the usual asymptotic series of $K_{\nu}$ and
carry out its commutation through the $n$-series, as explained in
\cite{erJMP}. Thus, we arrive at
\[
\sum_{n=-\infty}^{\infty}
\left( \pi n \over a \right)^{s-1/2} K_{s-1/2} (2\pi a n)=
\]
\beq
{1 \over 2}\Gamma\left( s -{1 \over 2} \right) a^{-2s+1}
+\pi^{s-1/2} a^{-s} \sum_{n=0}^{\infty}
{\Gamma(s+n) \over \Gamma(s-n) n!} {1 \over (4 \pi a)^n}
{\rm Li}_{n-s+1}(e^{-2\pi a}) .
\label{LiS}\eeq
where ${\rm Li}$ stands for the polylogarithm function
\beq {\rm Li}_s(z) \equiv \sum_{n=1}^{\infty} {z^n \over n^s}. \eeq
\req{LiS} is valid when $s>0$, which is satisfied here  since $s=(D-1)/2$
and we are taking $D>1$. Otherwise, extra terms arising from the
commutation
procedure can appear. Notice, too, that the $n$-series in \req{LiS}
is truncated for positive integer $s$ (odd $D$) due to the pole
of $\Gamma(s-n)$ downstairs.

%As for the first term in \req{Vgen2}, we are in conditions to use
%\req{GaAs}.
Using  \req{GaAs} for the first term in \req{Vgen2} together with the above
we have
%Gathering all these elements, one comes to
\beq
\begin{array}{ll}
\ds V_m(\theta)=
-{ \pi^{1 \over 2} \over
(4 \pi)^{D+1 \over 2}\Lambda^{D \over 2} } e^{ -i{D-2 \over 4}\theta }&
\ds\left\{ {e^{-x} \over x}
\left[ 1 + O\left( 1 \over x \right) \right] \right. \\
&\ds +y^{D \over 2} \left[
\pi^{-{D \over 2}}
\left( x \over y \right)^{D-1 \over 4} e^{-2\pi\sqrt{x/y}} \right. \\
&\ds\left.\hspace{3em}+O\left(
\left( x \over y \right)^{D-5 \over 4} e^{-2\pi\sqrt{x/y}},
\dots,
\left( x \over y \right)^{D-1 \over 4} e^{-4\pi\sqrt{x/y}},
\dots
\right)
\right] \\
&\ds\left. -2 \sum_{n=1}^{\infty} {y \over \pi^2 n^2}
e^{ -{\pi^2 n^2 \over y} -x}
\left[ 1 + O\left( y \over \pi^2 n^2 \right) \right]
\right\} .
\end{array}
\label{VlRlm}\eeq
%In the infinite-$R$ or flat-space limit, \req{VlRlm} reduces to the
%surviving part after setting
The infinite $R$ or topologically trivial  limit is $y=0$ and in this limit:
%$$y=0.$$
%The leading term of that contribution is
\beq
V_m(\theta)\approx
-{ \sqrt{\pi} \over (4 \pi)^{D+1 \over 2} \Lambda^{{D \over 2}+1} m^2 }
\ e^{\ds -\Lambda m^2 \cos{\theta \over 2}
-i \left[
{D \over 4} \theta + \Lambda m^2 \sin{\theta \over 2}
\right]
},
\label{Vplat2}\eeq
%\ie a potential of the same type as \req{Vplat}, already studied and
%shown in Fig. 1.
This is similar to \req{Vplat} and has already been discussed.

\subsection{Large radius and small mass
%$R \gg 1, m \ll 1$
$(|y|\ll 1, |x|\ll 1)$}

Since these conditions are not enough to fix the magnitude of
$|x|/|y|=(mR)^2$  we shall suppose
\beq \left\vert {x \over y} \right\vert \ll 1, \label{xoyss1} \eeq
The opposite case has already been discussed in the
previous subsection.

Fortunately  there is an identity linking our Bessel
function series and a particular zeta function, namely
\beq
E(s;a)\equiv\sum_{n=-\infty}^{\infty}( n^2+a^2 )^{-s}
={2 \sqrt{\pi} \over \Gamma(s)}\sum_{n=-\infty}^{\infty}
\left( \pi n \over a \right)^{s-1/2} K_{s-1/2} (2\pi a n),
\label{linkEK}\eeq
(for its derivation, see \eg \cite{erJMP})
which will be helpful in the evaluation of the potential \req{Vgen2} for
small values of $a\equiv\sqrt{x/y}$. It must also be born in mind
that in the end we  take $s=-(D-1)/2, D=2,3,\dots$.
In fact, for our purpose it will
be sufficient to know of an expansion of $E(s;a)$ in powers of
$a$, with due consideration of these values of $s$.
A similar problem was addressed in \cite{erIJMP} (see also \cite{EORBZ}),
and here we just quote the (adequately modified) result:
\[ \Gamma(s) E(s;a)= \]
\beq \left\{ \begin{array}{ll}
\ds\Gamma(s) a^{-2s}
-\sqrt{\pi}\Gamma\left( s-{1 \over 2} \right) a^{1-2s} \
\Theta\left( {1 \over 2} -s \right) & \\
\ds+2\sum_{n=0}^{\infty}(-1)^n{ \Gamma(s+n) \over n! }
\zeta(2n+2s)a^{2n}, &
\begin{array}{lll}s&\ne&{1\over 2}-m, \\ s&\ne&-m, \end{array} \\
\vspace{0.5cm} \\
\ds2\sum_{n=0}^{-s-1}(-1)^n{ \Gamma(s+n) \over n! }
\zeta(2n+2s)a^{2n}, &s=-m, \\
\vspace{0.5cm} \\
\ds\Gamma(s) a^{-2s}
+{(-1)^{1/2-s} \sqrt{\pi} a^{1-2s}
\over \left( {1 \over 2} -s \right)! }
\left[ \psi\left( 1 \over 2 \right)+ \psi\left( {1 \over 2} -s \right)
-\ln a^2 +2 \gamma \right] & \\
\ds+2\sum_{n=0 \atop n \ne 1/2 -s}^{\infty}(-1)^n{ \Gamma(s+n) \over n!
} \zeta(2n+2s)a^{2n}, &s={1\over 2}-m,
\end{array}\right.
\eeq
where $m=0,1,2,\dots$.
%As they stand,
The coefficients of these series
%yield undeterminations
%of the type infinity times zero, which are to be solved by
%application of
are to be understood with the aid of
the Riemann zeta function reflexion formula
\beq \Gamma\left( z \over 2 \right) \zeta(z)=
\pi^{z-1/2} \Gamma\left( 1-z \over 2 \right) \zeta(1-z). \eeq
%thus obtaining finite values.
In any of the above cases, the result found will be represented
by the series
\beq
\Gamma(s) E(s;a)=
c_{-2s}^{(-s)} a^{-2s} +c_{1-2s}^{(-s)} a^{1-2s}
+ c_{(L)}^{(-s)} a^{1-2s} \ln a^2
+\sum_{n \ge 0} c_n^{(-s)} a^{2n}.
\label{cPo}\eeq
Special values when $s$ is half an odd negative integer or a negative
integer $s$ are listed in Table \ref{Tb2}.

\begin{table}[htb]
\begin{center}
\begin{tabular}{rl}
$s$&power series for $\ds{1 \over \sqrt{\pi}}\Gamma(s)E(s; a)$ \\
\hline \vs
$-{1\over 2}$&$\ds {1 \over 3}-2a
+a^2\left[ \psi\left(1 \over 2 \right) + \gamma -\ln a^2 \right]
+{\zeta(3) \over 2}a^4 -{\zeta(5) \over 4}a^6 +{5 \zeta(7) \over 32}a^8
+O(a^{10})$ \\ \vs
$-1$&$\ds {\zeta(3) \over \pi^{5/2}}$ \\ \vs
$-{3\over 2}$&$\ds {1 \over 45}-{1\over 3}a^2+{4 \over 3}a^3
+{1 \over 2}a^4
\left[ \psi\left(1 \over 2 \right) +1+\gamma -\ln a^2 \right]
-{\zeta(3) \over 6}a^6 + {\zeta(5) \over 80} a^8 +O(a^{10})$ \\ \vs
$-2$&$\ds {3 \zeta(5) \over 2 \pi^{9/2}}
-{\zeta(3) \over \pi^{5/2}} a^2$ \\ \vs
$-{5\over 2}$&$\ds -{2 \over 945}+{1 \over 45}a^2+{1 \over 6}a^4
+{4 \over 15}a^5-{1 \over 6}a^6
\left[ \psi\left(1 \over 2 \right) +{3 \over 2} +\gamma -\ln a^2 \right]
+{\zeta(3) \over 24}a^8 +O(a^{10})$ \\  \vs
$-3$&$\ds {15 \zeta(7) \over 4 \pi^{13/2}}
-{3 \zeta(5) \over 2 \pi^{9/2}}a^2 +{\zeta(3) \over 2 \pi^{5/2}}a^4$. \\
\end{tabular}
\end{center}
\caption{$\pi^{-1/2}\Gamma(s)E(s; a)$ as a power series in $a$}
\label{Tb2}\end{table}

%Thus, once the particular form of \req{cPo} has been obtained, we
%use it to replace
We now replace
the Bessel series in \req{Vgen} with the appropriate form above, getting
\beq
\begin{array}{ll}
\ds V_m(\theta)=
-{ \pi^{1 \over 2} e^{ -i{D-2 \over 4}\theta } \over
(4 \pi)^{D+1 \over 2}\Lambda^{D \over 2} }&
\ds\left\{
b_{D \over 2}^{({D \over 2})} x^{D \over 2}
+ b_{(L)}^{({D \over 2})} x^{D \over 2} \ln x
+\sum_{n\ge 0} b_n^{({D \over 2})} x^n \right. \\
&\ds+{ y^{D \over 2} \over \sqrt{\pi} }
\left[
c_{D-1}^{({D-1 \over 2})} \left( x \over y \right)^{D-1 \over 2}
+c_D^{({D-1 \over 2})} \left( x \over y \right)^{D \over 2}
+c_{(L)}^{({D-1 \over 2})} \left( x \over y \right)^{D \over 2}
\ln \left( x \over y \right)
\right. \\
&\ds\hspace{3em} \left.
+\sum_{n \ge 0} c_n^{({D-1 \over 2})} \left( x \over y \right)^n
\right] \\
&\ds\left. -2 \sum_{n=1}^{\infty} {y \over \pi^2 n^2}
e^{ -{\pi^2 n^2 \over y} -x}
\left[ 1 + O\left( y \over \pi^2 n^2 \right) \right]
\right\}
\end{array}
\label{Vsxsxoy}\eeq
The simplest case to be considered is the topologically trivial  limit
$y=0$
when the main contribution comes from the leading term
$b_0$ given by \req{valb0},
%which gives rise to
\beq
V_m(\theta)\approx
-{2 \sqrt{\pi} \over (D-1) (4 \pi)^{D+1 \over 2} \Lambda^{D \over 2}}
\ e^{\ds -i{D-2 \over 4}\theta } .
\label{Vy0l}\eeq
When several fermions and bosons exist, \req{Vy0l} is
multiplied by $-(n_F-n_B)$.
As may be expected, this coincides with the induced potential
corresponding to massless fields in a no-ncompatified spacetime,
and has been  considered in \cite{gPLB,gcPre}. Ruling out the special
case $n_F=n_B$ or $D=2$ (which make $V$ independent of
$\theta$), the only solution which satisfies \req{requi} is
$n_F>n_B$ and $\theta=\pm\pi, D=4$.
%This has been interpreted by
%stating that, in
Thus, under these  conditions, a Lorentzian signature is singled
out by the dynamics only in $D=4$.

%The next thing to be considered is the
Deviations of this result due to mass corrections have been analysed in
ref.\cite{gcPre} in which all dimensions were considered non-compact.
It is therefore of interest to study the modifications induced by
compactification of various dimensions.
This requires the inclusion of $y$ contributions in
\req{Vsxsxoy}.

In order to be able to handle (order by
order) the resulting combined series, we will have to make a further
assumption on the relative size of $|x|$ and $|y|$. A simple hypothesis
consistent with \req{xoyss1} is
\beq O( |x| )=O( |y|^2 ). \label{TuningHyp} \eeq
With this hypothesis  we next consider the ensuing
modifications on the most interesting cases found in \cite{gcPre},

\subsubsection{$D=4$}

{}From \req{Vsxsxoy}, and using Tables \ref{Tb1} and \ref{Tb2}, we read
off the induced effective potential for a boson, up to orders
of $|x|^3$ or equivalent:
\beq
\begin{array}{ll}
\ds V_m(\theta)=-{ 1 \over 2 (4 \pi)^2 \Lambda^2 }&\ds\left\{
{1 \over 2} e^{-i {\theta \over 2}} -|x| \right. \\
&\ds+\left[
{ |y|^2 \over 45 }-{1 \over 3}|x||y|+{4 \over 3}|x|^{3/2} |y|^{1/2}
+{1 \over 2}\left( {3 \over 2}-\gamma \right) |x|^2
-{ |x|^2 \over 2 } \left( \ln |x| + i {\theta \over 2} \right)
\right. \\
&\ds\hspace{1em} \left. +{1 \over 2}|x|^2\left(
\psi\left( 1 \over 2 \right) +1 + \gamma - \ln {|x| \over |y|}
\right)
-{\zeta(3) \over 6} { |x|^3 \over |y| }
+{\zeta(5) \over 80} { |x|^4 \over |y|^2 } \right]
e^{i {\theta \over 2}} \\
&\ds \left. +{|x|^3 \over 6} e^{i\theta} + O( |x|^{7/2}, \dots )
\right\}
\end{array} .
\label{VsssD4}\eeq
Imposing the stationarity condition
\beq
\left.
{\partial \over \partial\theta }\imp V_m(\theta)
\right\vert_{\theta=\bar\theta}
=0
\eeq
on  \req{VsssD4}, we find
\[
\left[ 1
-{ 2 \over 45 }|y|^2+{2 \over 3}|x||y|-{8 \over 3}|x|^{3/2} |y|^{1/2}
+|x|^2 \left(
-{3 \over 2}-\psi\left( 1 \over 2 \right)+\ln{ |x|^2 \over y }
\right)
+{\zeta(3) \over 3} { |x|^3 \over |y| }
+{\zeta(5) \over 40} { |x|^4 \over |y|^2 } \right]
\cos {\bar\theta \over 2}
\]
\beq
-{|x|^2 \over 2}\bar\theta \sin {\bar\theta \over 2}
+{2 \over 3} |x|^3 \cos\bar\theta + O( |x|^{7/2}, \dots ) =0.
\eeq
We assume a solution around $\theta=\pm\pi$ of the type
\beq \bar\theta= \pm\pi + 2\varepsilon, \eeq
and find the approximate solution
%for $\varepsilon$, thus obtaining
\beq
\bar\theta \approx
\pm\pi\left[ 1-|x|^2 \left( 1 + {2 \over 45}|y|^2 \right) \right] .
\label{gcD4corrected}\eeq
The results so far apply to the case where only one boson is present.
We will now generalise them
%into a general form valid when several
%bosons and
%fermions, with different masses, exist.
to include bosons of mass $m_B$ and fermions of mass $m_F$.
For that purpose we introduce
the notations
\beq
\begin{array}{lll}
\Delta n&=& n_F-n_B, \\
\Delta|x|^n&=&\ds \sum_F |x_F|^n -\sum_B |x_B|^n,
\
\left( \left\vert x_{B \atop F} \right\vert  \equiv
\Lambda m_{B \atop F} \right), \\
\Delta(|x|^n \ln |x|) &=&\ds
\sum_F |x_F|^n \ln |x_F| - \sum_B |x_B|^n \ln |x_B|.
\end{array}
\eeq
%In order to exhibit the dependence of $V$ on $x$ (which contains
%all the mass information) we symbolically write
$$V(\theta)=V[x(\theta), y(\theta)]\equiv V(x,y).$$
%As already commented, when several fields are present, the whole
%potential becomes
%a sum of the $V$'s associated to each, weighted with factors $+1$ ($-1$)
%for bosonic (fermionic) fields ---for reasons explained in
%\cite{gPLB,gcPre}. It is not difficult to realize that,
In order to
generalise the previous expressions, it is enough to make the following
replacements
\beq
\begin{array}{lll}
V(x,y)&\to&\ds \sum_B V(x_B,y) - \sum_F V(x_F, y) \\
\mbox{$y$-independent terms}&\to&\mbox{$y$-independent terms $\cdot
(-\Delta n)$ } \\
|x|^n&\to&-\Delta|x|^n \\
|x|^n \ln |x| &\to&-\Delta(|x|^n \ln |x|) .
\end{array}
\label{ReplGenV}\eeq
With these changes  the solution becomes
\beq
\bar\theta \approx
\pm\pi\left[ 1-{\Delta|x|^2\over \Delta n}
\left( 1 + {2 \over 45}|y|^2 \right) \right] .
\eeq
The $y$-independent correction is the one found in \cite{gcPre},
while the $|y|^2$-term provides a first measure of the deviation due to
a large-$R$ effect.
Making also the hypothesis
\beq {\Delta|x|^2\over \Delta n} <0, \eeq
we find
that the approximate $\bar\theta$ falls outside
of the domain considered, \ie
$$ \bar\theta \not\in [ -\pi, \pi ]. $$

Now, we turn our attention to the real part of \req{VsssD4}, which,
up to the same order as $\imp V_m$, reads
\beq
\begin{array}{ll}
\ds \rep V_m(\theta)=-{ 1 \over 2 (4 \pi)^2 \Lambda^2 }&
\ds\left\{ \left[ {1 \over 2}
+{ |y|^2 \over 45 }-{1 \over 3}|x||y|+{4 \over 3}|x|^{3/2} |y|^{1/2}
+{1 \over 2}\left( {3 \over 2}-\gamma \right) |x|^2
-{ |x|^2 \over 2 } \ln |x| \right. \right. \nn
&\ds\hspace{1em} \left. +{1 \over 2}|x|^2\left(
\psi\left( 1 \over 2 \right) +1 + \gamma - \ln {|x| \over |y|}
\right)
-{\zeta(3) \over 6} { |x|^3 \over |y| }
+{\zeta(5) \over 80} { |x|^4 \over |y|^2 } \right]
\cos{\theta \over 2} \\
&\ds\left.
-{|x|^2 \over 4}\theta \cos {\theta \over 2}
+{|x|^3 \over 6} \sin\theta + O( |x|^{7/2}, \dots )
\right\} .
\end{array}
\eeq
Its leading term
is proportional to $\cos \theta/2$ so generalising to include all masses we
have
\beq
\rep V(\theta)={ 1 \over 2 (4 \pi)^2 \Lambda^2 }
\left[ \Delta n \left( {1 \over 2} + {|y|^2 \over 45} \right)
\cos{\theta\over 2}
+\Delta |x| + O( \Delta |x|^{3/2} ) \right] .
\eeq
Notice that, if  $n_F>n_B$, one has $\Delta n >0$ and the leading
term is
%thus a nonnegative sinusoidal
for $\theta \in [-\pi, \pi]$. Obviously,
when restricted to this interval, this curve has two absolute minima
at $\theta=\pm\pi$.
The first
correction is $\theta$-independent,
%and has the only effect of shifting
%up or down this sinusoidal, leaving
and leaves the location of these minima
unchanged.

\subsubsection{$D=6$}
For $D=6$, eq. \req{Vsxsxoy} and the information
in Tables \ref{Tb1} and \ref{Tb2} give
\beq
\begin{array}{ll}
\ds V_m(\theta)=-{ 1 \over 2 (4 \pi)^3 \Lambda^3 }&\ds\left\{
{1 \over 3} e^{-i \theta } - {1 \over 2}|x|e^{-i {\theta \over 2} }
+{|x|^2 \over 2} \right. \\
&\ds+\left[
 -{2 \over 945 }|y|^3-{1 \over 45}|x||y|^2+{1 \over 6}|x|^2|y|
+{4 \over 15}|x|^{5/2} |y|^{1/2} \right. \\
&\ds\hspace{1em}\left. +{1 \over 6} |x|^3 \left(
-{ 10 \over 3 } + \ln {|x|^2 \over |y|} - \psi\left( 1 \over 2 \right)
+ i {\theta \over 2}
\right)
\right] e^{-i {\theta \over 2}} \\
&\ds\left.+ O( |x|^{7/2}, \dots )
\right\},
\label{VsssD6}\end{array}
\eeq
and the stationarity condition on its imaginary part reads
\[
-{4 \over 3}\cos\bar\theta
\]
\[
+\left[
|x| -{ 4  \over 945 }|y|^3 +{2 \over 45} |x||y|^2
+{1 \over 3}|x|^2|y| +{8 \over 15}|x|^{5/2}|y|^{1/2}
+{1 \over 3}|x|^3 \left(
-{17 \over 6}-\psi\left( 1 \over 2 \right)+\ln{ |x|^2 \over |y| }
\right)
\right]
\cos {\bar\theta \over 2}
\]
\beq
-{|x|^3 \over 6}\bar\theta \sin {\bar\theta \over 2}
+ O( |x|^{7/2}, \dots ) =0.
\label{statD6}\eeq

We now make the replacements \req{ReplGenV} on $V_m$. Further, following
\cite{gcPre} we suppose
\beq \Delta n =0, \label{Dn0}\eeq
and realize that condition \req{statD6} becomes
\[
-\left[
\Delta |x| +{2 \over 45} \Delta |x||y|^2
+{1 \over 3} \Delta |x|^2|y| +{8 \over 15} \Delta |x|^{5/2}|y|^{1/2}
\right.
\]
\[
\left.
+{1 \over 3} \Delta |x|^3 \left(
-{17 \over 6}-\psi\left( 1 \over 2 \right)-\ln |y|
\right)
+{2 \over 3} \Delta( |x|^3 \ln |x| )
\right]
\cos {\bar\theta \over 2}
\]
\beq
+{ \Delta |x|^3 \over 6}\bar\theta \sin {\bar\theta \over 2}
+ O( \Delta |x|^{7/2}, \dots ) =0.
\eeq
Solving as before, we find an approximate solution to be
\beq
\bar\theta \approx
\pm\pi\left[ 1-{1 \over 3}{\Delta|x|^3\over \Delta |x|}
\left( 1 - {2 \over 45}|y|^2 \right) \right] ,
\label{gcD6corrected}\eeq
where, again, the $y$-dependent part provides the first
large $R$ correction to the result in \cite{gcPre}.
One might think that this
correction vanishes for $|y|=\sqrt{45/2}$, but this value
lies outside the
%looks too large to be in the range of
 present {\it small}-$|y|$ approximation.
So, taking $|y|<\sqrt{45/2}$, everything depends on
$\Delta|x|^3/\Delta |x|$. Assuming
\beq {\Delta|x|^3\over \Delta |x|} <0, \eeq
we are lead to the same situation as in the previous case, namely,
that the stationary points are outside of the domain $[ -\pi, \pi ]$.
%Then, we look at
The real part of \req{VsssD6} is
\beq
\begin{array}{ll}
\ds \rep V_m(\theta)=-{ 1 \over 2 (4 \pi)^3 \Lambda^3 }&
\ds\left\{ {1 \over 3}\cos\theta +{|x|^2 \over 2} \right. \\
&\ds+ \left[
-{1 \over 2}|x|
-{2 \over 945}|y|^3 +{1 \over 45}|x||y|^2+{1 \over 6}|x|^2|y|
+{5 \over 15}|x|^{5/2} |y|^{1/2} \right. \\
&\ds \left. \hspace{1em} +{1 \over 6}|x|^3 \left(
-{10 \over 3} -\psi\left( 1 \over 2 \right) + \ln {|x|^2 \over |y|}
\right)
\right]
\cos{\theta \over 2} \\
&\ds\left.
+{|x|^3 \over 12} \theta \sin{\theta \over 2} + O( |x|^{7/2}, \dots )
\right\} .
\end{array}
\eeq
Generalizing  by means of \req{ReplGenV}, and recalling the assumption
\req{Dn0}, we find its first nonvanishing terms to be
\beq
\rep V(\theta)={ 1 \over 2 (4 \pi)^3 \Lambda^3 }
\left[  {1 \over 2} \Delta |x| \cos{\theta\over 2}
+{1 \over 2}\Delta |x|^2 + O( \Delta |x|^2 |y|, \Delta |x|^3, \dots )
\right] .
\eeq
If
\beq \Delta|x| >0, \eeq
we get the same as in the $D=4$ case, \ie the leading term is a
nonnegative sinusoidal, vanishing at $\theta=\pm\pi$, multiplied by a
positive constant, and the next-to-leading correction is
$\theta$-independent. Therefore, the conclusion is the same. Within
this approximation, the minima of $\rep V$ on $[-\pi, \pi]$ are at
$\theta=\pm\pi$.

%Making the same considerations as before, we realize
%that this is the the preferred solution as regards stability.

\section{Closing Remarks}

Regarding signature as  a dynamical quantity, we have examined its
influence
on the one-loop effective potential.
Using Grensite's criteria for a stable potential we have analysed the
dependence of a preferred signature on
both the inclusion of massive fields and a compactified background.
We have argued that a  flat metric with a Lorentzian signature on
${\bf R}^{D-1} \times {\bf S}^1$ may arise in a number of distinct ways.

We extend the  solutions found previously in $D=4$ and $D=6$ by examining
their stability under the influence of compactification and massive fields.
For suitably  large masses the nature of potential
 as a function of the compactification radius
(\req{Vplat} and \req{Vplat2}) indicates that the preferred  signature is
insensitive to compactification.
If the  requirements on the solution are  relaxed  slightly by
accepting  approximate coincidence of minima and stationary points,
we get similar results
in 4- and 6-dimensional universes with small $m$ and close to the
topologically trivial  limit $R \to \infty$.
Assuming
that $\Lambda$ and $m^2R^4$ are comparable (\req{TuningHyp})
deviations from the previous cases have been evaluated in first
approximation.
Eqs. \req{gcD4corrected} and \req{gcD6corrected} provide a first
measure of the separation from Greensite's stationary points
($\bar\theta$) for small $m>0$ due to weak compactification effects in
${\bf R}^3 \times {\bf S}^1$ and ${\bf R}^5 \times {\bf S}^1$,
respectively. The value of $\bar\theta=(\pm\pi)$
found in ref. \cite{gcPre} receives now a multiplicative correction of
the form $(1+\mbox{constant} |y|^2)$, where $|y|^2=\Lambda^2/R^4$.
The radius-independent parts are in agreement with ref. \cite{gcPre},
which dealt with topologically trivial spaces.

In view of these results one may   argue that, for our present
universe with  $D=4$,  1-loop quantum
fluctuations away from a flat metric with a Lorentzian signature are
largely suppressed according to the hypotheses that we have adopted.
Clearly this result may be modified by the inclusion of gravity.
%An analysis of the influence of strong curvature on the effective
%potential of the dynamical Wick field will be given elsewhere.

\appendix\section{Appendix: curvature effects in De Sitter spaces}

The same sort of analysis can be easily applied to the study of
curvature effects in the presence of the dynamical Wick angle.
As an example, we consider the potential for a massless scalar
field in an
${\bf R}^{D-N} \times {\bf S}^N$
spacetime, where ${\bf S}^N$ is an $N$ dimensional De Sitter space of
radius $a$. As usual $\xi$ will denote the coupling with the scalar
curvature $R$, so that our operator is now $\Box+\xi R$. After
integrating over the continuous momentum components, the effective
potential reads
\beq
V( \theta )=
- { [f(\theta )]^{-{D-N \over 2}+1} \over
2 (4 \pi)^{D-N \over 2} {\rm Vol}( {\bf S}^N ) }
\sum_{n=0}^{\infty} \left( \bar\lambda_n y \right)^{-{D-N \over 2}}
\Gamma\left( -{D-N \over 2}, \bar\lambda_n y \right),
\eeq
where $\bar\lambda_n$ and $d_n$ are the dimensionless eigenvalues and
their degeneracies
\beq
\left\{
\begin{array}{lll}
\lambda_n&=&\ds{1 \over a^2}\bar\lambda_n
={1 \over a^2}[ n^2+n(N-1)+ \xi R a^2 ], \\
d_n&=&\ds (2n+N-1){ (n+N-2)! \over n! (N-1)! },
\end{array}
\right.
\eeq
and
\beq
\begin{array}{lll}
f( \theta )&=& e^{i {\theta \over 2}}, \\
y&=& f(\theta) {\Lambda \over a^2} .
\end{array}
\eeq

First, we look at the large-$|y|$ limit, which implies small radius or
large curvature. If $\xi \ne 0$, the leading term of the resulting
expansion has the form
\beq
V( \theta ) \approx
-{1 \over
2 (4 \pi)^{D-N \over 2} {\rm Vol}( {\bf S}^N ) \xi R \Lambda} \
e^{\ds -i {D-N \over 4}\theta
-\xi R \Lambda\left(\cos{\theta\over 2} + i \sin{\theta\over 2}\right)
} .
\eeq
Since we suppose $R$ to be large, this gives a plateau-like potential,
which indicates the above commented lack of preference for any
signature. On the other hand, carefully repeating the calculation for
$\xi=0$ one obtains
\beq
V( \theta ) \approx
-{1 \over
(D-N) (4 \pi \Lambda)^{D-N \over 2} {\rm Vol}( {\bf S}^N ) } \
e^{\ds -i {D-N-2 \over 4}\theta } .
\eeq
In the light of the previous discussions, it is clear that this produces
Lorentzian signature when $D=N+4$, \ie for spacetimes of the type
${\bf R}^4 \times {\bf S}^N$, which includes the Kaluza-Klein-style
solution for $N=1$.

Of course, taking the opposite case, or small-$|y|$, brings nothing
new. This corresponds to the flat-space limit and, once more, gives a
main contribution to $V(\theta)$ of the form constant
$\ds \cdot e^{-i {D-2 \over 4}\theta }$, which reproduces Greensite's
original set-up, with its ensuing Lorentzian solution for $D=4$.

\hskip2cm

\ni{\Large \bf Acknowledgements}

This research was supported in part by the European Union under the Human
Capital and Mobility programme.
The authors are grateful to Emili Elizalde for helpful discussions.
S.D.O. would like to thank members of the Deptartment ECM for their kind
hospitality.

\newpage

\ni{\Large \bf Figure Caption}

\ni{\bf Figure 1}.
Changing nature of a (rescaled) potential of the
type $\ds v(\theta)=e^{\ds -B \cos{\theta \over 2}
+i\left( A \theta+ B\sin{\theta \over 2} \right) }$,
for $-\pi \le \theta \le \pi$.
Curves drawn in solid and dashed line represent $\rep v(\theta)$
and $\imp v(\theta)$, respectively.
%with $a\equiv{D-1 \over 4}$, $b\equiv{\Lambda \over 4 R^2}$
The plots shown correspond to
fixed $A=0.75$ and to different values of $B$:
(a) $B=1$, (b) $ B=10$.
In (b) the formation of a wide plateau around the origin
%for large values of $B$
is already noticeable. The width of this plateau can be
seen to increase as $B$ grows.
Changes in the value of $A$ do not produce significant alterations on
this varying behaviour.

\end{document}